\documentclass[12pt]{article}
\usepackage[utf8]{inputenc}
\usepackage[margin=1in]{geometry}

\usepackage{hyperref}
\usepackage{graphicx,subcaption}
\usepackage{amsmath, amsthm, amssymb}
\usepackage{bm}
\usepackage[ruled,vlined]{algorithm2e}
\SetKwInput{KwInput}{Input}
\SetKwInput{KwOutput}{Output}
\usepackage{multirow}
\usepackage{makecell}
\usepackage{adjustbox}
\usepackage{xcolor}
\usepackage{authblk}

\DeclareMathOperator*{\argmax}{arg\,max}

\usepackage[sort, numbers, compress]{natbib}
\def\TheoremsNumberedThrough{%
\theoremstyle{TH}%

\newtheorem{proposition}{Proposition}

\theoremstyle{EX}

\newtheorem{definition}{Definition}

}
\usepackage{setspace}

\TheoremsNumberedThrough

\begin{document}

\title{Empirical Characteristics of Affordable Care Act Risk Transfer Payments}

\author{Grace Guan\thanks{Department of Management Science and Engineering, Stanford University, Stanford, USA. G.G. is supported by the National Science Foundation Graduate Research Fellowship under Grant No. 1656518. Email: {\tt gzguan@stanford.edu.}} ~~and Mark Braverman\thanks{Department of Computer Science, Princeton University, Princeton, USA. M.B. is supported in part by the NSF Alan T. Waterman Award, Grant No. 1933331, a Packard Fellowship in Science and Engineering, and the Simons Collaboration on Algorithms and Geometry. Any opinion, findings, and conclusions or recommendations expressed in this material are those of the authors and do not necessarily reflect the views of the National Science Foundation.}}

\date{March 25, 2022}

\setcounter{Maxaffil}{0}
\renewcommand\Affilfont{\itshape\small}

\maketitle

\begin{abstract}
Under the Affordable Care Act (ACA), insurers cannot engage in medical underwriting and thus face perverse incentives to engage in risk selection and discourage low-value patients from enrolling in their plans. One ACA program intended to reduce the effects of risk selection is risk adjustment. Under a risk adjustment program, insurers with less healthy enrollees receive risk transfer payments from insurers with healthier enrollees. Our goal is to understand the elements driving risk transfers. First, the distribution of risk transfers should be based on random health shocks, which are unpredictable events that negatively affect health status. Second, risk transfers could be influenced by factors unique to each insurer, such as certain plans attracting certain patients, the extent to which carriers engage in risk selection, and the degree of upcoding. We create a publicly available dataset using Centers for Medicare and Medicaid Services data that includes insurer risk transfer payments, costs, and premiums for the 2014-2017 benefit years. Using this dataset, we find that the empirical distribution of risk transfer payments is not consistent with the lack of risk selection as measured by the ACA risk transfer formula. Over all states included in our dataset, at least 60\% of the volume of transfers cannot be accounted for by a purely normal model. Because we find that it is very unlikely that risk transfer payments are caused solely by random shocks that reflect health events of the population, our work raises important questions about the causes of heterogeneity in risk transfers.
\end{abstract}

\clearpage

\onehalfspacing

\section{Introduction}

The 2010 Patient Protection and Affordable Care Act (ACA) worked towards closely related policy goals: reducing the cost of and increasing access to healthcare insurance. Yet from 2010 to 2019, national health spending in the United States increased from \$2.60 trillion to \$3.8 trillion, and from 17.3 percent of the gross domestic product (GDP) to 17.7 percent of the GDP \citep{antos_aca_2020, cms_national_2019}. In a review of the healthcare operations management literature, \cite{dai_om_2020} ask the question: “Are insurance companies simply victims of providers or physician power or are they complicit in this failure [of cost reduction in American healthcare], due to perverse incentives that are unintended consequences of well-intentioned health policy?”

Risk adjustment was one policy intended to mitigate insurers’ aversion to enrolling higher-cost patients, thus increasing access to insurance. To reward insurers for expanding coverage at lower premiums, the ACA risk adjustment program provides risk transfer payments  (also called ``risk transfers'' or ``transfers''), as computed by a risk transfer formula, to insurers to eliminate the effects of risk selection on premiums \citep{pope_risk_2014}. Because transfers are in part based on patient diagnoses, the risk adjustment program creates financial incentives for insurers to collect greater reimbursement. Indeed, in the past, risk adjustment programs as implemented in Medicare Advantage have led to unintended consequences of “upcoding” patient diagnoses to maximize the profitability of enrolling sicker-than-average patients \citep{geruso_upcoding_2019}, and enrolling higher-value individuals who cost less than their reimbursements from risk transfer payments \citep{brown_how_2014}. We aim to see whether there are similar unintended consequences of the ACA risk adjustment program.

In this paper, we create a dataset using publicly available insurer financial data from 2014-2017. We use this dataset to model whether risk transfer payments of the ACA risk adjustment program truly offset the effects of risk selection on plan costs. We find that the empirical distribution of risk transfer payments is not consistent with the lack of risk selection as measured by the ACA risk transfer formula. 

\subsection{Background}
Prior to the ACA, an insurer could profit by enrolling more low-risk, healthy individuals who would lower the average cost of that insurer's risk pool, while simultaneously turning away or charging higher premiums to high-risk, sicker, or more expensive individuals \citep{bertko_what_2016}. The ACA eliminated this type of behavior through guaranteed issue and community rating \citep{cox_explaining_2016, claxton_pre-existing_2016}. Guaranteed issue prevents insurers from turning away consumers with pre-existing conditions. Community rating prohibits insurers from charging consumers different prices within defined geographical regions based on factors such as age, gender, or health status. Together, guaranteed issue and community rating effectively stop insurers from participating in medical underwriting, or using the consumer's medical history in determining whether or not to provide insurance, and if so, what premium to charge. Consequently, insurers who must now enroll pricier individuals may find themselves operating at a loss, regardless of whether this occurred by chance, reputation, or plan features. 

In such a market environment without a risk adjustment program, insurers would engage in risk selection \citep{cox_explaining_2016}. Specifically, insurers face a significant incentive not only to find a way of getting rid of their sickest patients (e.g. by treating them poorly to get them to switch insurers), but also to try to attract healthier patients by designing plans unattractive to people with expensive health conditions (e.g., in what benefits they cover or through their drug formularies). Risk selection decreases efficiency in the insurance market because insurers compete with each other to attract a certain type of individual (healthy individuals) and do not focus on providing value to their enrollees by driving down costs. In an ideal system, insurers would be competing based on plan attributes and quality of care rather than cherry-picking enrollees. 

A risk adjustment program that makes insurers indifferent to the prior health statuses of individuals in their risk pool could incentivize insurers to compete through productive activities (e.g., better care management to reduce costly adverse outcomes) or neutral activities (e.g., negotiating better rates from providers). A risk adjustment program accomplishes this through risk transfer payments compensating insurers with higher-risk enrollees through payments from insurers with lower-risk enrollees. Risk transfer payments are intended to make up for the extra healthcare costs associated with high-risk individuals; it also allows insurers to charge similar, competitive premiums so as to not discourage healthy patients from enrolling \citep{pope_risk_2014}.

However, a sub-optimal risk adjustment program could also give insurers an incentive to engage in undesired behavior. For example, if an insurer learns that lung cancer patients on average made the insurer more money via risk transfer payments than the insurer's average costs for lung cancer patients, insurers would face significant incentive to enroll more lung cancer patients. When a risk adjustment program is in play, insurers now have significant incentive to get rid of their lowest-value patients, that is, the most expensive patients when accounting for their costs less their transfer payment reimbursements. Thus, it is important to investigate the extent to which risk transfer payments account for risk selection to better understand the implications of this policy for future policy and providers.

The ACA’s risk adjustment program is described in a series of three articles published in 2014, when risk adjustment, reinsurance, and risk corridors were implemented. In these articles, Kautter et al. and Pope et al. introduce a risk adjustment methodology to complement the ACA, which includes a risk adjustment model and a risk transfer formula \citep{kautter_hhs-hcc_2014, kautter_affordable_2014, pope_risk_2014}. As with any other risk adjustment strategy, the ACA risk adjustment formula and model encourage insurers to enroll unhealthy people, consequently stabilizing premiums and costs \citep{kautter_hhs-hcc_2014}. The risk transfer payments to insurers who enroll sicker people (who have positive transfers) are taken from insurers who enroll healthier individuals (who thus have negative transfers) \citep{pope_risk_2014}. Within each state, these transfer payments sum to zero \citep{pope_risk_2014}. These payments are calculated per plan by averaging each patient’s “risk score,” which is an expectation of how much a patient should cost relative to a healthy patient, calculated using demographic factors and objective pre-existing conditions \citep{kautter_affordable_2014}. A more detailed background on risk adjustment that motivates this work is given in Appendix \ref{Appendix:risk_adjustment_background}.

Our work is related to incentives in insurance programs, for which there is much literature \citep{bastani_evidence_2019, brown_how_2014,braverman_data-driven_2021}, whose root is in the principal-agent problem \citep{plambeck_performance-based_2000}. Previous work has also focused on analyzing risk adjustment under Medicare Advantage and other programs \citep{brown_how_2014, einav_selection_2013,  geruso_upcoding_2019, layton_risk_2016, glazer_optimal_2000, newhouse_steps_2012} and the ACA exchanges \citep{adida_bundled_2017, dickstein_impact_2015, frean_premium_2017, einav_market_nodate, zink_fair_2020, tebaldi_estimating_nodate, mcguire_improving_2021}. To the best of our knowledge, our paper is the the first to identify evidence of of risk selection in the ACA markets.

\subsection{Main Contributions}
We create a publicly available dataset, consolidating data available online from the Centers for Medicare \& Medicaid Services (CMS) website. This dataset includes insurer risk transfer payments, costs, premiums, and size for the 2014-2017 benefit years. We then develop a model to estimate ACA risk transfer payments if they were to fully offset the effects of risk selection on plan costs. Calibrating the model with small-group market insurer-level costs and size data from our dataset, we estimate the amount of money changing hands not due to differences in risk scores. We find that the empirical distribution of risk transfer payments is not consistent with the lack of risk selection as measured by the ACA risk transfer formula.

\section{Data}

We obtained insurer financial data from the Centers for Medicare and Medicaid Services (CMS) website. We combined two data sources over the 2014-2017 benefit years to create a dataset, which we made publicly available. 

Our dataset combined two sources of data: (1) more than 20,000 Medical Loss Ratio “Insurer Report” Excel files, one for each insurer operating in a given state, which contain reported values of risk transfers, member months, premiums, and costs; (2) “Summary Report” PDF files for each benefit year, which contain the actual risk adjustment transfer payments for each insurer in each state for each benefit year. Each source by itself was insufficient for analysis, (1) due to non-finalized risk transfers and (2) due lack of insurer financials. Our
dataset documentation is given in Appendix \ref{appendix:dataset_documentation}.

The final dataset contained all insurers who reported values over all marketplaces. In our analysis, we considered only the small-group market, and we defined each state as competitive or not competitive.

\subsection{Dataset Creation}

Our dataset and its documentation are available at LINK TO BE PROVIDED AFTER PEER REVIEW.

Both the ``Insurer Report'' Excel files\footnote{\href{https://www.cms.gov/apps/mlr/mlr-search.aspx}{https://www.cms.gov/apps/mlr/mlr-search.aspx}} and the ``Summary Report'' PDF files\footnote{\href{https://www.cms.gov/CCIIO/Programs-and-Initiatives/Premium-Stabilization-Programs/index.html}{https://www.cms.gov/CCIIO/Programs-and-Initiatives/Premium-Stabilization-Programs/index.html}} were publicly available online and were downloaded from the CMS website. The data was available for the 2014-2018 benefit years. We excluded the 2018 benefit year data from our analysis because it was missing over 10\% of insurers.

For each ``Insurer Report" Excel file, only the first 3 sheets were relevant to our analysis. The first sheet contained insurer information, including company name, Health Insurance and Oversight System identification code (HIOS ID), non-profit status, and state. The second sheet contained Premiums, Claims, Taxes, Healthcare Quality Improvement Expenses, Non-Claims Costs, Income from Uninsured Plans' Fees, Amount Insured (Member Months, Number of Covered Lives, etc.), and Net Investment. The third sheet contained detailed information about Premiums and Claims. We manually decrypted any encrypted files. These files alone were inadequate, because risk transfers were not final and did not sum to zero within each state.

The ``Summary Report'' PDF file contained values for HIOS ID, company name, company state, reinsurance payment (up to the 2016 benefit year), and individual and small group risk adjustment payments. Each benefit year had two PDF files, one for estimated data that matched the ``Insurer Report'' Excel files as well as the ``Public Use'' Excel files, and one with the updated final values of the payments that ended up being made for each benefit year. We extracted data from the PDF files containing the final payment values by first copying the data into a plain text file, then parsing the values within the plain text file and adding them into a table. Since company names often had line breaks, we parsed each raw text file by whitespace and determined what each field was through reading every word until finding a value that represented a payment. Of the words that we skipped, the last word represented the state, and previous words represented the company name. These files alone were inadequate because they did not contain member month, premiums, or costs data.

We combined the ``Insurer Report'' and ``Summary Report'' files into one CSV file for each benefit year in 2014-2017. To generate the final dataset, we matched the ``Summary Report'' PDF Files that contained the final risk adjustment values to the ``Public Use'' Excel file through an outer merge on HIOS ID. In our final dataset, each row represented an insurer operating in a given state from an ``Insurer Report'' Excel file. The columns include insurer characteristics (e.g., name, non-profit status, state) and insurer financials (e.g., costs, premiums, member months).  

\subsubsection{CMS ``Public Use'' Files}
The CMS website also provides ``Public Use'' Excel files from 2011-2018 at \href{https://www.cms.gov/CCIIO/Resources/Data-Resources/mlr.html}{https://www.cms.gov/CCIIO/Resources/Data-Resources/mlr.html} for each benefit year, which supposedly contain all data from every company for each benefit year. These files were inadequate for our analysis, because they did not contain company identification and had significant missing attributes (e.g., insurer company names and HIOS IDs). Due to the missing attributes, we were unable to match the data with the ``Summary Report'' PDF files that contained each benefit year's final risk transfer payments. For example, of the companies listed in this dataset, of the five with the smallest transfers per member month, we were only able to match three with the``Summary Report'' PDF file.

\subsection{Data Sample Selection}
\label{competitive}

In our analysis, we used looked at companies with more than $2,000$ member months per year (approximately $166$ enrolled individuals per year) in the individual group market, and no missing values for the fields of Premiums, Costs, Transfers, Reinsurance, or Member Months.

\subsubsection{Choice of Marketplace}

Three different marketplaces are present in the ACA exchanges: the individual marketplace, the small group marketplace, and the large group marketplace. In the individual marketplace, individuals can autonomously choose their plan. In contrast, group marketplaces enroll businesses as a whole, which includes all of their employees. The small group marketplace provides insurance for “small employers,” which are defined as employers with less than 100 full-time employees. Accordingly, the large group marketplace provides insurance for “large employers,” who have more than 100 full-time employees. Risk adjustment was implemented for the ACA individual and small group marketplaces. We focused our analysis on the small group marketplace because insurers have less choice over patients than if the patients were to enroll by themselves in the individual market.  

\subsubsection{Determining ``Competitive'' States}

To exclude states where one insurer covers a majority of the market, we defined states as competitive or not competitive. We computed Herfindahl-Hirschman Indexes (HHI), the sum of squares of each insurer's market share, for each state's small group market. A state was defined as competitive if its HHI was below average in the 2015 benefit year. These states, listed in alphabetical order, are: Arizona, Arkansas, California, Colorado, Connecticut, Florida, Georgia, Hawaii, Maine, Maryland, Michigan, Minnesota, Missouri, Nebraska, Nevada, New Jersey, New Mexico, New York, Ohio, Oregon, Pennsylvania, Virginia, Washington, Wisconsin, and Utah.

\section{Model}

In the small-group ACA market, the distribution of risk transfers under the risk adjustment program should be based on \textbf{health shocks}, which are unpredictable illnesses that affect health status negatively. Due to the largely unforeseeable nature of medical cost utilization, we make the assumption that these shocks should be relatively normally distributed around a mean of zero with a variance that represents some variability in patient sickness. We determine whether the empirical distribution of risk transfers has a non-normally-distributed, linearly-scaling, component in addition to this normally-distributed health shocks component. Such a component could consist of ``gaming’’ or genuine sorting. Gaming would occur if an insurer actively attempts to attract a certain more profitable type of patient. Similarly, genuine sorting occurs if certain insurers are better at coding (transcribing patient diagnoses into bills) or more attractive to sick patients (e.g. a certain insurer has a famous name brand, as compared to a startup).

Finding such a non-normally distributed component is significant because the ultimate goal of a risk adjustment scheme is to compensate health insurance plans for differences in the health of their enrollees. The health mix of each plan is represented by the average of each patient’s individual risk score, which reflects health shocks and thus should be randomly distributed in the small group market where insurers cannot cherry-pick enrollees. Transfer payments ensure that plan premiums reflect differences in plan factors, as opposed to variance in enrollees’ health statuses or other factors, so there would be no incentive to turn patients away and no punishment for enrolling sicker patients. If other factors are in play in determining risk transfer payments, the effectiveness of the policy is greatly reduced because risk transfer payments will not fully compensate insurers for the effects of risk selection on premiums. 

\subsection{Formulation}

Our model considered a set $I = \{1, ..., k\}$ of insurers operating during a given benefit year. Each insurer in our analysis is an insurance company operating in a given state. Thus, the same insurance company operating in two different states would count in our analysis as two distinct insurers. In our analysis, we excluded states where one insurer covers a majority of the market, marketplaces other than the small group market, and insurers who had fewer than $2,000$ member months, or approximately $166$ enrollees. We present results for three states that are relatively large and have competitive marketplaces (New York, California, and Wisconsin), those three marketplaces combined, all competitive states, two randomly divided selections of states, and all states. Section \ref{competitive}  presents further details on sample selection.

Each insurer $i\in I$ receives or pays risk transfers $T_i$, where $T_i > 0$ indicates that the insurer is making money from transfers, and $T_i < 0$ indicates that the insurer is losing money from transfers. Within each state, $T_i$ sums to $0$. The $T_i$ are distributed as the random variable $T$. Each insurer faces costs $C_i > 0$ (distributed as the random variable $C$) and enrolls $n_i$ patients. All transfers and costs were normalized to be in average-state 2017 dollars.

Underlying these insurers' enrollees is the true variance of expected sickness of patients under all insurers, which we denote $b^2$.

As in the literature (\cite{einav_selection_2013,ho_health_2021}), we assume that risk transfers per patient are log-normally distributed. Since each insurer enrolls thousands of patients, by the Central Limit Theorem, the normalized sum of an insurer's patient-level transfers (denoted by $T_i / n_i$) should tend towards a normal distribution.

\begin{proposition}
\label{variance_scales_by_sqrt_enrollees}
For insurer $i$ with number of enrollees $n_i$, the random health shock component of the risk transfer payment $T_i$ is given by $T_i \sim \mathcal{N}(0,b^2) * \sqrt{n_i}$.
\end{proposition}

For our model of $T_i$, we add a linear cost term that scales with $n_i$, because this term is due to any other factor unrelated to random health shocks. This ``drift'' in patient sickness would be the same for every patient of an insurer. Though there could be sublinear drifts as well from plan variation, linear drifts are the most important from a policy perspective. Thus, our model is given as:
\begin{equation}
   T_i \sim \mathcal{N}(0,b^2) * \sqrt{n_i} + C_i * n_i.
   \label{eq:T_definition}
\end{equation}

Each $T_i$ as in Equation \ref{eq:T_definition} has an interpretation as the summation of two independent terms, with the first term $\mathcal{N}(0,b^2) * \sqrt{n_i}$ representing true random health shocks due to fluctuations in patient sickness, and the second term $C_i * n_i$ reflecting genuine variations in patient mix, gaming, upcoding, or other things. 

Denote $\overline{T_i}$ as an insurer's transfers normalized by the square root of their enrollees ($T_i / \sqrt{n_i}$). To be able to identify the existence of the cost term in our model, we consider the distribution of all empirical $\overline{T_i}$'s, $\overline{T}$:
\begin{equation}
    \overline{T} \sim \mathcal{N}(0,b^2) + C * \sqrt{n}.
    \label{eq:hypothesis}
\end{equation}

\section{Empirical Strategies for Identifying Non-Random Factors Influencing Transfers}
\label{section:estimation}

We first estimate variation in patient sickness by finding points of significant difference between empirical risk transfers and simulations based on Equation \ref{eq:T_definition}. Based on these estimations, we calculate the likelihood that the realized, empirical transfers follow our model. Further tests that make weaker assumptions and are less powerful, (i.e., no existence of a linear cost term) are given in Appendix \ref{Appendix:additional_tests}.

\subsection{Estimating Variation in Patient Sickness}

The true variance of expected sickness of patients under all insurers $b^2$ cannot be accurately estimated because the transfer payments to insurers will have more variance than the true costs insurers pay due to the risk scores given by the transfer formula and upcoding. We denote the empirical estimation of $b^2$ as $\beta^2$. Thus, assuming that underlying patient cost and transfer formula noise are independent,
\begin{equation}
    b^2 = Var(\text{underlying patient cost}) + Var(\text{transfer formula
noise} \mid \text{true patient cost}).
\end{equation}

So, we lower bound $b^2$ by an estimate $\beta^2$ which drops the transfer formula noise. Therefore, $\beta^2 < b^2$.

We estimate $\beta^2$ in three ways: comparing our empirical data to a half-normal distribution, simulations of risk transfers, and simulations of risk transfers at a fixed percentile.

\subsubsection{Comparison to a half-normal distribution} 

We now consider a half-normal distribution of the form $|\mathcal{N}(0,a^2)|$, so we can use properties of stochastic dominance in estimating the national variance in patient sickness. An example of this is graphically shown in Figure \ref{fig:estimating_beta_example} in Appendix \ref{Appendix:supplemental_results_2}. 

\begin{definition}
\label{stochastic_dominance}
For a random variable $Y$, we denote $F_{Y}(v) = P(Y\leq v)$ as the cumulative distribution function of $Y$. Given two random variables $Y$ and $Z$, $Y$ \textit{stochastically dominates} $Z$ over a range $A$ if for all $v\in A$, $F_Y(v) \leq F_Z(v)$. Denote $Y$ stochastically dominating $Z$ as $Y \geq_{st} Z$.
\end{definition}

Recall the formulation of transfers given in Equation \ref{eq:hypothesis}, which has a strictly positive $C*\sqrt{n}$ term. From Definition \ref{stochastic_dominance}, there should exist some range such that
\begin{equation}
     |\mathcal{N}(0,\beta^2)|\geq_{st} |\overline{T}|.
\end{equation}

Now, we consider some basic properties of stochastic dominance.

\begin{proposition}
\label{stochastic_dominance_proposition}
Let $Y\sim |\mathcal{N}(0,y)|$ and $Z\sim |\mathcal{N}(0,z)|$. If $0\leq y<z$ then $Z\geq_{st}Y$.
\end{proposition}

As we are still interested in the relationship to the unknown true variance of expected sickness of patients, because $\beta^2<b^2$, then by Proposition \ref{stochastic_dominance_proposition}:

\begin{equation}
 |\mathcal{N}(0,b^2)| \geq_{st} |\mathcal{N}(0,\beta^2)|  \geq_{st} |\overline{T}|
\end{equation}

To estimate $\beta$, there cannot be significant deviations between $|\overline{T}|$ and $|\mathcal{N}(0,\beta^2)|$. We first define the point where the greatest deviation occurs, as well as its length, as follows.

\begin{definition}
\label{point_of_maximum_difference}
Given a half-normal random variable with variance $a^2$, $|\mathcal{N}(0,a^2)| $, and $|\overline{T}|$, define the \textit{point of maximum difference} as 
\begin{equation}
    v_{a}  = \argmax_v \left[ F_{|\overline{T}|}(v) - F_{|\mathcal{N}(0,a^2)|}(v) \right].
\end{equation}

Correspondingly, the \textit{length of maximum difference} is
\begin{equation}
    M_{a} = F_{|\overline{T}|}(v_{a}) - F_{|\mathcal{N}(0,a^2)|}(v_{a}).
\end{equation}

\end{definition}

Given the point and length of maximum difference between the two CDFs, we can quantify whether it is significant as follows.

\begin{definition}
\label{point_of_dominance} Let $K$ follow the one-sided Kolmogorov distribution, and $k$ be the number of insurers in the market. Let the significance of our test be $\alpha = 0.05$. Define $c_{\alpha}$ as $P(K \leq c_{\alpha}) = 1-\alpha$. If $M_a\sqrt{k} > c_{\alpha}$, then there exists a statistically significant \textit{point of dominance} between $|\mathcal{N}(0,a^2)|$ and $|\overline{T}|$. 
\end{definition}

\begin{proposition}
\label{point_of_dominance_remains}
If there is a point of dominance $v_{\beta}$ when $|\overline{T}|$ is compared to $|\mathcal{N}(0,\beta^2)|$, for any $b>\beta$, $v_{\beta}$ will remain a point of dominance when $|\overline{T}|$ is compared to $|\mathcal{N}(0,b^2)|$. 
\end{proposition}

We report the maximum value of $\beta$ for which there is no point of maximum difference between $|\mathcal{N}(0,\beta^2)|$ and $|\overline{T}|$.

\subsubsection{Comparison to simulations of risk transfers} In addition to comparing $|\overline{T}|$ to a half-normal distribution, we use information about insurers' market fraction to simulate their yearly transfer payments.

For each insurer $i$, we simulated costs $C'_i$ with variance proportional to both the number of enrollees $n_i$ and overall variance in patient sickness $\beta^2$:
\begin{equation}
    C'_i \sim \mathcal{N}(0, \beta^2 \cdot n_i).
\end{equation}

Denote simulated insurer transfers by $T'$. As we assume that costs scale linearly by enrollees, we simulated insurers' transfers $T'_i$ by subtracting insurer $i$'s fraction of the market $\left(\frac{n_i}{\sum_{j=1}^{k}n_j}\right)$ multiplied by total simulated market costs, from that insurer's simulated costs:
\begin{align}
    T'_i &= C'_i - \left( \sum_{j=1}^{k}C'_j \right) \left(\frac{n_i}{\sum_{j=1}^{k}n_j}\right) \\
    &= C'_i \left(\frac{\sum_{j=1, j \neq i}^{k}n_j}{\sum_{j=1}^{k}n_j}\right) - \frac{n_i}{\sum_{h=1}^{k}n_h} \sum_{j=1, j \neq i}^{k} C'_j \\
    &\sim \mathcal{N}(0, \beta^2 \cdot a_i)
\label{eq:xtilde}
\end{align}

Where $a_i$ is defined as
\begin{equation}a_i = \left( n_i \left(\frac{\sum_{j=1, j \neq i}^{k}n_j}{\sum_{j=1}^{k}n_j}\right)^2 + \left(\frac{n_i}{\sum_{h=1}^{k}n_h}\right)^2 \sum_{j=1, j \neq i}^{k} n_j \right)~~.
\label{eq:aiequation}
\end{equation}

A summary of the simulation is given in Algorithm \ref{alg:simulated_transfers}. We generate the normalized simulated series, $\overline{T'_i}$, $15$ times to ensure the sample size is sufficiently large. Then, we calculate the largest $\beta$ for which there is no point of dominance between the CDFs of $|\overline{T'_i}|$ and $|\overline{T}|$. We run the overall simulation $5$ times and report the average value of $\beta$. 

\begin{algorithm}[htbp]
\SetAlgoLined
 $J = 15$ = number of iterations\;
 \For{$j = 1, 2, ..., J$}{
     \For{each state-year pair $(s, y)$}{
          \For{each insurer $i$ in $(s, y)$}{
           Sample $T'_i$ as in Equation \ref{eq:xtilde}\;
           } {
           Normalize $T'_i$ to be in average 2017 dollars\;}
      }
          Return series $\{\overline{T'_i} = T'_i / \sqrt{n_i}, \forall i \in I\}$\;
 }
 Return $J$ samples of series of $\{\overline{T'_i}, \forall i \in I\}$\;
 \caption{Simulation of insurer risk-transfers given $\beta$}
 \label{alg:simulated_transfers}
\end{algorithm}

\subsubsection{Comparison to simulations of risk transfers at a fixed percentile}

One drawback of the previous two approaches is that they are dependent on the Kolmogorov distribution. Therefore, given a fixed percentile, we aim to create simulations to find a $\beta$ where the empirical distribution does not significantly differ from our simulations. An example of this is graphically shown in Figure \ref{fig:estimating_beta_example_2} in Appendix \ref{Appendix:supplemental_results_2}. 

Given a fixed percentile $p$, where
\begin{equation}
    P(|\overline{T_i}| < v_{empirical}) = P(|\overline{T'_i}| < v_{simulation}) = p,
\end{equation}

we find the minimum $\beta$ such that for every $10^5$ generated samples of $|\overline{T'_i}|$ where
\begin{equation}
    P(|\overline{T_i}| < v_{empirical}) = p,
\end{equation}
at most 1\% of them have $v_{simulation} < v_{empirical}$.

That is, at percentile $p$, $v_{empirical}$ is in the smallest 1\% of simulated risk transfer CDFs. We found the smallest $\beta$ for which a point of dominance existed for a significant proportion of simulated distributions.

\subsection{Identifying Non-Random Factors Influencing Transfers}
\label{section:identifying_factors}
After calculating the largest possible $\beta$ for which there is no point of dominance between the CDFs of $|\overline{T}|$ and $|\mathcal{N}(0,\beta^2)|$ or the simulations, by Proposition \ref{point_of_dominance_remains}, we know that for any $\beta' > \beta$, a point of dominance between the CDFs exists.
Then, we report the proportion of realized transfers $|\overline{T_i}|$ that lie outside 2 standard deviations of the mean, $P(|\overline{T_i}|>2\beta)$. We also report the probability (``p-value'') that there exist this many or more realized transfers above $2\beta$ in our empirical distribution given a simulation generated using variance $\beta^2$. One such occurrence has probability $P(|\overline{T_i}|>2\beta)= 1- erf(\sqrt{2}) \approx 0.0455$. If this probability is greater than $\alpha$, then for no value of $\beta$ will Equation \ref{eq:hypothesis} hold. That is, if $\beta$ were truly smaller than it is, then the proportion of realized transfers with magnitude above $2\beta$ only increases, and the probability that there exist this many or more realized transfers with magnitude above $2\beta$ decreases. 

\section{Proportion of Empirical Risk Transfers Based on Other Factors}

To understand what fraction of a risk transfer payment comes from factors other than health shocks, we create a lower bound by modeling how much money would be transfered between insurers in expectation if the values of $\beta$ estimated in Section \ref{section:estimation} were the true standard deviation of a single person-year expense. Because the estimated $\beta$ is an upper bound, this estimation method is conservative.

Given the estimated values of $\beta$, we first calculate a formula for expectation of absolute transfers. We simulate each insurer's transfers $|T'_i|$ as in Equation \ref{eq:xtilde}, so by linearity of expectation:

\begin{equation}\text{Expectation of Absolute 2017 Risk Transfers} = E\left(\sum_{i\in I} |T'_i|\right) = \frac{\sqrt{2}\beta}{\sqrt{\pi}}\sum_{i\in I}\sqrt{a_i},\label{eq:ExpectationT}\end{equation}

where $a_i$ is given in Equation \ref{eq:aiequation}. This expectation is in average state 2017 dollars. We compare the estimation in Equation \ref{eq:ExpectationT} to the empirical sum of absolute transfers in 2017, 

\begin{equation}
    \text{Realized Absolute 2017 Risk Transfers} = \sum_{i\in I} |T_i|.
    \label{eq:RealT}
\end{equation}

We expect the realized amount of money changing hands to be more than the expectation of the amount of money changing hands based on simulated $\beta$. We calculate the ratio $f$ by which the sum of empirical absolute transfers exceeds the sum of simulated absolute transfers, 
\begin{equation}
    f = \frac{\sum_{i\in I} |T_i|}{E\left(\sum_{i\in I} |T'_i|\right)}.
\end{equation}

It follows that the fraction of money that changes hands due to reasons that are not normal shocks in patient sickness is given by $1 - 1/f$.

\section{Results}

We report results over three states with competitive markets, New York, California, and Wisconsin, these three states combined, and competitive states. For results on additional samples, see Table \ref{tab:beta_estimation_results_Appendix} in Appendix 
\ref{Appendix:supplemental_results}. 

\subsection{Estimation of national variation in patient sickness}

\begin{table}[htbp]
\centering
\begin{tabular}{|c|c|c|c|c|c|c|c|}
\hline
\multicolumn{2}{|c|}{\textbf{Comparison to}} & \multicolumn{3}{c|}{\textbf{Half-normal distribution}} & \multicolumn{3}{c|}{\textbf{Simulation described in Alg. \ref{alg:simulated_transfers}}} \\ \hline
\textbf{State(s)} & \textbf{$k$} & \textbf{$\beta$} & \textbf{$P(|\overline{T_i}| >2\beta)$} & \textbf{p-value} & \textbf{$\beta$} & \textbf{$P(|\overline{T_i}| >2\beta)$} & \textbf{p-value} \\ \hline
NY only & 79 & 81,100 & 0.177 & $<10^{-4}$ & 101,320 & 0.152 & $<10^{-4}$ \\ \hline
CA only & 61 & 83,400 & 0.115 & $0.006$ & 86,260 & 0.115 & $0.006$ \\ \hline
WI only & 104 & 13,800 & 0.183 & $<10^{-6}$ & 13,300 & 0.202 & $<10^{-6}$ \\ \hline
NY, CA, and WI & 244 & 28,000 & 0.250 & $<10^{-6}$ & 29,380 & 0.234 & $<10^{-6}$ \\ \hline
Competitive & 1,233 & 23,800 & 0.172 & $<10^{-6}$ & 22,380 & 0.181 & $<10^{-6}$ \\ \hline
\end{tabular}

\caption{Estimated $\beta$ from comparing the Empirical $|\overline{T}|$ against $|\mathcal{N}(0,\beta^2)|$ and simulated $|\overline{T'}|$. Corresponding proportion of realized transfers greater than $2\beta$, and the p-value of this occurring.}
\label{tab:beta_estimation_results}
\end{table}

\begin{table}[htbp]
\centering
\begin{tabular}{|c|c|c|c|c|c|}
\hline
\textbf{State(s)} & $k$ & $p$ & $\beta$ & $P(|\overline{T_i}| > 2\beta)$ & \textbf{p-value} \\ \hline
Competitive & 1,233 & 10 & 19,987 & 0.213 & $<10^{-6}$ \\ \hline
All & 1,966 & 10 & 14,861 & 0.241 & $<10^{-6}$ \\ \hline
\end{tabular}
\vspace{0.3cm}
\caption{Estimated $\beta$ for which any larger $\beta$ would have a point of dominance for 99\% of simulated distributions using the fixed percentiles to generate the values for $\beta$.}
\label{tab:q10testresults}
\end{table}

Tables \ref{tab:beta_estimation_results} and \ref{tab:q10testresults} present the maximum estimated $\beta$ such that there is no point of dominance between the CDFs of the empirical $|\overline{T}|$ and fixed distribution $|\mathcal{N}(0,\beta^2)|$ or simulated $|\overline{T'}|$, the fraction of realized transfers that have absolute value greater than $2\beta$, and the probability that there exist this many or more realized transfers above $2\beta$ in our empirical distribution given a simulation generated using only random health shocks with variance $\beta^2$ (``p-value''). 

\subsection{Identifying Non-Random Factors Influencing Transfers}

No matter the $\beta$, group of states, or method used to generate $\beta$, the probability that the realized distribution of transfers has so many outliers is well below $0.01$. More than $10\%$ of insurers' transfer payments fall outside $2\beta$ when comparing to a half-normal distribution or the simulation described in \ref{alg:simulated_transfers} (Table \ref{tab:beta_estimation_results}).  Over any range of states that includes New York, since the absolute value of the maximum risk transfer payment in that state was  $\$6.45\times 10^5$, there exists one risk-transfer payment with magnitude above $6\beta$ compared to the rest of the New York sample alone. This transfer is also larger than $30\beta$ when compared to a sample that combines other states. As any event outside 4 standard deviations of the mean has a probability less than $10^{-4}$ of occurring, the existence of this outlier provides a strong signal for the existence of other factors influencing insurer's risk transfers.

\subsection{Proportion of Empirical Risk Transfers Based on Other Factors}

Table \ref{tab:ratio_simulated_expected} compares the empirical amount of money changing hands to the average sum of simulated amount of money changing hands (over $10^5$ trials). In line with our hypothesis, we observe that the magnitude of the empirical transfers exceed the magnitude of the simulated transfers by a factor $f$ between 1.32 and 3.15. In larger samples, over 60\% of transfers cannot be explained by normal random shocks. Although we normalized for differences in state costs, we expected larger sample sizes to have a higher proportion of transfers that are not explainable by random shocks due to variation between states in insurer proportion of market. Thus, a relatively low-cost, competitive state such as Wisconsin may only have less than 30\% of its transfer volume accounted for by non-random factors, whereas, a higher-cost, competitive state with many outliers such as New York has around 50\% of its transfer volume accounted for by non-random factors. When adding states with less competitive exchanges to calculate the ratio by which simulated transfers exceed empirical transfers over all states that we have data for, the proportion of money changing hands not accountable for by our simulations increases to 0.68 from 0.64.

\begin{table}[htbp]
\centering
\begin{tabular}{|c|c|c|c|c|}
\hline
 & \multicolumn{2}{c|}{$E(\sum_{i\in I} |T'_i|)$*} & \multicolumn{2}{c|}{$E(\sum_{i\in I} |T'_i|)$**} \\ \hline
\multicolumn{1}{|c|}{\textbf{State(s)}} & $f$ & $1-1/f$ & $f$ & $1-1/f$ \\ \hline
\multicolumn{1}{|c|}{NY only} & 1.76 & 0.43 & 2.00 & 0.50 \\ \hline
\multicolumn{1}{|c|}{CA only} & 1.99 & 0.50 & 1.84 & 0.46 \\ \hline
\multicolumn{1}{|c|}{WI only} & 1.44 & 0.30 & 1.32 & 0.24 \\ \hline
\multicolumn{1}{|c|}{Competitive} & 2.79 & 0.64 & 2.78 & 0.64 \\ \hline
\multicolumn{1}{|c|}{Random States A} & 2.80 & 0.64 & 3.01 & 0.67 \\ \hline
\multicolumn{1}{|c|}{Random States B} & 2.91 & 0.66 & 2.86 & 0.65 \\ \hline
\multicolumn{1}{|c|}{All} & 3.15 & 0.68 & 3.13 & 0.68 \\ \hline
\end{tabular}\\
\caption{The ratio by which the simulated transfers exceed the empirical transfers and the ratio of money changing hands not due to random shocks in patient sickness in 2017 dollars.}
\label{tab:ratio_simulated_expected}
* from Alg. \ref{alg:simulated_transfers}
** from percentile simulations
\end{table}

\section{Discussion}

We create a consolidated dataset on insurer size and financials, including final risk transfers, using CMS data from the 2014-2017 benefit years. Using this dataset, we find that it is very unlikely that transfer payments are caused solely by random health shocks, so there exist significant non-random factors that threaten the efficacy of the risk adjustment program. Our work was motivated by the potential for opportunistic gaming or innocent genuine variation in patient mix that might undermine the current implementation of risk adjustment under the ACA. Our results raise questions about the causes of hetereogeneity in risk transfers. 

Assuming that risk transfers in the small group market should be based on normally distributed health shocks, we simulated the distribution of transfer payments, we compared the realized distribution of risk transfers to samples from a normal distribution as well as simulations of insurers' risk transfer payments. In both cases, no standard deviation of expected patient sickness allowed us to both (1) ensure the realized distribution of risk transfers did not have a point of significant difference from our normal distribution or simulations and (2) account for the large number of insurers whose risk transfers' magnitudes were above 2 standard deviations of the mean.

We lower bounded the proportion of risk transfers that could not be accounted for by our simulations. In larger samples, such as insurers over the 25 most competitive states in 2015, taking a random selection of half of the states, or $1,966$ insurers over all of the states we had data for, at least $60\%$ of transfer volume could not be accounted for with our simulations.

One limitation of our work was that the insurer level data did not allow us to tease out the difference between genuine variation in patient mix and risk selection in insurers' response to risk adjustment. However, the identification of some ``drift'' in patient sickness, that would be the same for every patient of an insurer, is an important first step to understanding what factors drive risk transfer payments. Future work should investigate what factors play into risk transfers that have magnitude substantially far from the mean.

\subsection{Implications for Policy}
In this work, we have demonstrated how ACA risk transfers from 2014-2017 in the small group market are not adequately compensating insurers for variance in health status, as could be measured by the sum of ``true random health shocks’’. This research points towards a strong ``linear drift’’ component of transfers that could be due to gaming, upcoding, genuine variations in patient mix, or other factors. 

To discourage further risk selection that lead to shifts in patient mix and gaming, the ACA should look towards factors that were used in the the MA program to reduce favorable selection \citep{newhouse_steps_2012}. These factors included using enrollee diagnosis to predict risk scores and a prohibition on monthly disenrollment by beneficiaries. First, disenrollment could be prevented through mandated enrollment periods or re-implementing the individual mandate of the ACA. Second, while it is impossible to perfectly predict individual patient costs, the ACA risk transfer formula already uses enrollee diagnosis to predict medical expenditure risk similar to MA \citep{kautter_hhs-hcc_2014, newhouse_steps_2012}. 

More research is needed to gain insights for policy that could address insurers' varying patient mixes and differing proclivity towards gaming. For example, future work could stratify our dataset into those insurers into that carry a national brand name and those insurers that appear to be more startup-like to investigate shifts in patient mix. This stratification could reveal how specific classes of insurers' transfer payments compare to their premiums and costs, and could bring to light whether sicker patients decide to choose brand name insurers over a startup. Similarly, if the outliers that prevent insurers' distributions from being normal are largely due to catastrophic illnesses, further work could explore if programs such as reinsurance and risk corridors could mitigate the effects of these outliers. Reinsurance and risk corridors give insurers additional premium if certain enrollees have ``catastrophic costs'' or if medical costs are greater than a fixed percentage of target costs respectively, and these programs were phased out of the ACA after the 2016 benefit year \citep{layton_risk_2016}.

Although the ACA successfully preventing carriers from engaging in medical underwriting and made large strides towards its goals of reducing the cost of and increasing access to health insurance, there remain many unanswered questions with respect to the ACA’s affordability, market stability, and protections of pre-existing conditions. The potential for gaming and genuine variations in patient mix may lead to risk selection that could potentially undermine the effectiveness of the risk transfer program.

This work has shown that while risk transfers under the ACA are supposed to be linked to random health shocks, there in fact exist significant non-normal components to the risk transfer payments of the ACA. Differences in enrollee risk status will always exist in healthcare, so fair payment must be given to insurers who by chance enroll sicker than average individuals. If the risk adjustment scheme of the ACA improves, insurers will be able to focus on providing better care at lower costs, rather than competing on the basis of attracting the highest-value patients.

\clearpage

\bibliographystyle{vancouver}
\bibliography{references}

\clearpage
\appendix

\section*{Appendix}

\section{Proofs}
\label{Appendix:proofs}

\proof{Proof of Proposition \ref{variance_scales_by_sqrt_enrollees}.} 
First, consider $T_i/n_i$, which is normally distributed based on the CLT. Since each $n_i$ is large and the population average individual patient-level risk-transfer is $0$, the mean of $T_i/n_i$ is $0$. Therefore, the mean of $T_i$ is also $0$.

Now, consider $T_i$ as the sum of i.i.d. health shocks, because we assumed that it is impossible to select patients in the small group market. Let i.i.d. random variables $H_1, H_2, ..., H_{n_i}$ represent health shocks, each distributed as a random variable $H$ with variance $b^2$, so $$Var(T_i) = Var\left(\sum_{j=1}^{n_i}H_j\right) = \sum_{j=1}^{n_i}Var\left(H_j\right) = n_i Var(H) = n_ib^2.$$

We conclude $T_i \sim \mathcal{N}(0,b^2) * \sqrt{n_i}$.
\endproof

\proof{Proof of Proposition \ref{stochastic_dominance_proposition}.} Choose $x\in \mathbb{R}_{+}$. Then, $$F_Y(x) = P(-x\leq Y \leq x) = erf\left(\frac{x}{\sqrt{2}y}\right) \geq erf\left(\frac{x}{\sqrt{2}z}\right) = P(-x \leq Z \leq x) = F_Z(x).$$ Since the choice of $x$ was arbitrary, then $Z\geq_{st}Y$.
\endproof

\proof{Proof of Proposition \ref{point_of_dominance_remains}.} 
From Definition \ref{stochastic_dominance}, $F_{|\mathcal{N}(0,b^2)|}(v_{\beta}) \leq F_{|\mathcal{N}(0,\beta^2)|}(v_{\beta})$, so by Definition \ref{point_of_maximum_difference}, $M_b \geq M_{\beta}$. Thus, $M_b \sqrt{k} \geq M_\beta \sqrt{k} > c_\alpha$, and by Definition \ref{point_of_dominance}, $v_{\beta}$ is also a point of dominance between $|\mathcal{N}(0,b^2)|$ and $|\overline{T}|$.
\endproof

\section{A Background on Risk Adjustment}
\label{Appendix:risk_adjustment_background}

In this section, we provide motivation and additional background context for our work. We first explain the necessity of risk adjustment as a component of the ACA. We then survey the history of risk adjustment programs including Medicare Advantage and Medicare Part D in America, as well as programs in various other countries.

In theory, implementing a risk adjustment strategy is necessary for a competitive insurance market to operate without discrimination against individuals with pre-existing conditions. However, the ACA may face serious problems with risk selection in the future due to inherent flaws in risk adjustment that incentivize insurers to cherry-pick select types of consumers. Historical examples of risk adjustment leading to risk selection from Medicare Advantage, Medicare Part D, and in countries such as Belgium, Germany, Israel, the Netherlands and Switzerland, provide important motivation for this work, which investigates whether the ACA risk adjustment program adequately compensates insurers for variance in patient health status.

\subsection{Why Is Risk Adjustment Necessary?}

Prior to the enactment of the ACA, people with pre-existing conditions were charged higher premiums or denied insurance coverage \citep{bertko_what_2016}. Pre-existing conditions ranged from acute conditions such as acne to chronic conditions such as diabetes and cystic fibrosis for which there could be a lifetime of additional medical expenses \citep{bertko_what_2016}. According to an October 2019 analysis of the prevalence of pre-existing conditions by the Kaiser Family Foundation, nearly 54 million Americans under 65 currently have pre-existing conditions that would make them uninsurable in the individual market without the ACA \citep{claxton_pre-existing_2016}. These 54 million Americans represent 27\% of all adults under 65 \citep{claxton_pre-existing_2016}. Due to the ACA-mandated guaranteed issue provision, all ACA insurers operating in the individual market are now required to provide coverage to anyone who can pay their premium, regardless of pre-existing conditions. However, even with the mandate of guaranteed issue, insurers may not truly provide fair coverage to every enrollee in the individual market. 

To drive a more profitable business, insurers would prefer to avoid paying for enrollees that could potentially have annual medical costs in the tens or hundreds of thousands of dollars. \cite{bertko_what_2016} note that insurers can avoid these sicker-than-average people both through innocuous methods (e.g., ``aim advertising at healthy people who use gyms’’) and through more controversial methods (e.g., ``develop networks of physicians or hospitals that discourage access to care with, for example, fewer oncologists’’). In response, a risk adjustment strategy must be implemented to minimize insurers’ avoidance of people who are more likely to utilize care. Risk adjustment thus seeks to ``level the playing field’’ between insurers who by chance end up with larger amounts of sicker patients.

Fair coverage issues must also be addressed in the small group market, which covers small businesses up to 100 employees. Risk adjustment must be implemented in the small group market, because small employers may not have the resources to pool costs if one employee gets sick. In states where insurers consider individual employee claims and health status as factors in determining an employer's small group rating, a single employee illness can result in significant increases in premiums for the employer.

The fundamental theory underlying the ACA risk adjustment scheme is the same as in any other risk adjustment system. There exist a certain number of discrete health condition categories for which insurers with sicker-than-average enrollees will obtain extra payments beyond the premiums that they charge. However, unlike other systems, these ACA risk transfer payments do not come from the government, but rather from insurers who have healthier-than-average enrollees who pay into a ``risk adjustment pot’’ to make the payments. There is no doubt that some form of a risk adjustment strategy is needed to prevent discrimination against people with pre-existing conditions, the adverse selection death spiral, and market unraveling. To be clear, the ACA’s other provisions prevent discrimination by mandating insurers enroll any patient who applies, but absent a risk adjustment plan, insurers will continue to try to get rid of any low-value patients.

Risk adjustment as a policy directive has received mixed reviews from scholars, who analyze the program with various frameworks and different market assumptions. \cite{Geruso2016} analyzed trade-offs in a risk adjustment system, finding that risk adjustment causes a fundamental tradeoff between limiting overall costs on the one hand, and limiting the provision of care only to high-value customers (``cream-skimming’’). Introducing a simple reinsurance program (one-time payments for expensive enrollees) would help resolve this trade-off dilemma and lead to more efficient markets \citep{Geruso2016}. On the other hand, \cite{Layton2017} showed that even though risk adjustment causes health plan premiums to be based on costs that are not coded in the discrete health condition categories of the risk adjustment model, the welfare benefit of risk adjustment is over \$600 per person-year. 

The authors of the series of papers that introduced the ACA risk transfer formula and framework give theoretical examples for how risk adjustment under the ACA would work \citep{pope_risk_2014}.  The authors’ simplest example includes a market with two plans of equal coverage, Plan 1 and Plan 2, with enrollees of differing health status. Specifically, Plan 1's expected enrollees are sicker than those of Plan 2. Plan 1 charges a monthly premium of \$466.67 and Plan 2 charges a monthly premium of \$233.33, because premiums are set proportional to expected medical costs. However, with risk adjustment, Plan 2 would pay \$116.67 to Plan 1. Because both plans would adjust for the impact of risk transfers on their costs, they should both set premiums of \$350 in the expectation that Plan 1 enrolls a 33\% sicker population than the mean. The risk adjustment capitation payment thus makes these plans equivalent to consumers. Further examples provided by the authors illustrate the incentives insurers face to adjust premiums for plans of unequal coverage, plans with different enrollee age mixes and geographic costs, and plans of different tiers. The examples provide a foundation on which we understand the mechanism of risk transfers in theory.

\subsection{Risk Adjustment in the Real World}

To understand the potential impacts of the ACA’s design of risk adjustment, we must not only review the concepts of benefit design and risk adjustment in theory, but also review the impact of risk adjustment programs on risk selection in real-world implementation. Specifically, we review analyses performed on Medicare Advantage, on Medicare Part D, and on government risk-adjusted ``sickness funds’’ of various health systems in other countries.

\subsubsection{Medicare Advantage}

Medicare Advantage (MA) has been host to the largest risk adjustment effort to date in the US healthcare sector. The risk adjustment program of MA inspired the risk adjustment program of the ACA. A key difference between the MA and ACA risk adjustment programs is that ACA payments are zero-sum, in other words entirely paid for by insurers, whereas MA payments come from the government. Evidence from the history of MA illustrates that after risk adjustment was implemented, MA insurers no longer felt the need to avoid patients with conditions included in the risk transfer formula. In fact, MA insurers now try to select for these patients due to the higher risk-adjusted payments that they garner for the MA carrier hosting the plan they enroll in.

Since the 1980s, MA, formerly known as the Medicare private policies program and Medicare Part C, has allowed individuals to enroll in either a traditional, government-administered fee-for-service (FFS) program or in a privately-administered MA plan \citep{brown_how_2014}. At a minimum, MA plans must cover the benefits guaranteed by FFS programs but can also provide additional services. As of 2014, over 25\% of Medicare's 52 million enrollees chose the private MA option \citep{brown_how_2014}. As of 2019, this number is up to almost 35\% of Medicare's enrollees \citep{KFFMedicare}. The government pays the MA plan the cost of providing an enrollee their Medicare benefits through a capitation payment. In 2000, the Centers for Medicare \& Medicaid Services (CMS) began using inpatient diagnostic information for risk adjustment for MA \citep{bertko_what_2016}. Patient diagnoses from medical visits were added in 2004, so capitation payments became based on an individual's risk score, generated by a risk-adjustment formula that included more than 70 disease conditions \citep{bertko_what_2016, brown_how_2014}. 

Before risk adjustment was introduced for MA, MA plans had an incentive to over-emphasize the enrollment of healthy, low-cost individuals. After risk adjustment, however, plans no longer avoided patients with conditions included in the formula. In fact, plans now greatly appreciate the benefits of risk adjustment. By way of example, in late 2015, Aetna, one of the biggest U.S. insurers, began advertising to diabetics, since they believed the risk adjustment capitation payments would allow them to profit from enrolling more diabetics, even as they helped these diabetics manage care better and control costs \citep{bertko_what_2016}.

Although MA plans no longer avoided patients with conditions in the formula, they still engaged in various forms of patient selection. \cite{brown_how_2014} used a difference-in-differences model to show the occurrence of selection after the implementation of risk-adjustment. First, relative to enrollees in FFS, the risk scores of those joining MA increase after risk adjustment, so people enrolling in MA were on average sicker, but only as measured by the scores of the risk adjustment formula \citep{brown_how_2014}. However, the actual costs conditional on the risk score of those joining MA fell substantially after risk adjustment relative to enrollees in FFS, indicating that MA plans found individuals who were ``cheap for their risk score’’ \citep{brown_how_2014}. For example, the authors found that: ``pre-risk-adjustment, Hispanics were roughly \$1,200 cheaper on average than their (non-risk-adjusted) capitation payments; after risk adjustment, Hispanics with a history of congestive heart failure (one of the most common conditions included in the risk formula) are on average \$3,500 cheaper than their (risk-adjusted) capitation payments’’ \citep{brown_how_2014}. This finding confirms the authors’ predictions that MA plans enrolled individuals with higher scores but lower costs conditional on their scores.

\subsubsection{Medicare Part D} Medicare Part D is a prescription drug insurance program for seniors subsidized by the U.S. government. Similar to the ACA and MA, Medicare Part D also subsidizes insurers on the basis of enrollee diagnoses using a risk adjustment formula. Building off of \cite{brown_how_2014}, which studied Medicare Advantage, \cite{Carey2017} investigated the presence of technological change in the forms of new drug introduction and onset of generic competition on profitability of enrollee diagnoses. This work shows \textit{how} the design of risk adjustment enables unfavorable risk selection of profitable individuals \citep{Carey2017}. 

\cite{Carey2017} ultimately found that ``several highly effective new drugs for the diagnosis of multiple sclerosis (MS) entered after the risk adjustment system was calibrated, while major branded anti-hypertensives began to face generic competition. The typical plan receives \$334 per MS patient but spends \$1,185; meanwhile, the plan receives \$207 per enrollee with hypertension, but spends only \$147.'' Because of a miscalibration in the risk adjustment system, MS patients became unprofitable whereas hypertension patients became profitable -- an example as to how risk adjustment benefit design, here in Medicare Part D, can create incentives to select patients with certain diagnoses \citep{Carey2017}. As a result, \cite{Carey2017} suggests policy changes, specifically out-of-pocket cost regulations on certain therapeutic classes of drugs to decentivize insurers from selecting patients based on diagnosis alone.

\subsubsection{Risk Adjustment in Other Countries} 

\cite{Vandeven2007} analyzed the relationship between risk adjustment and risk selection in Belgium, Germany, Israel, the Netherlands, and Switzerland from 2000 to 2006. At the time, these countries had risk-bearing sickness funds (similar to insurers) with risk-adjusted premium subsidies so that consumers’ premiums to their sickness funds are not based on differences in risk profiles \citep{Vandeven2007}. These countries employ a range of risk adjustment factors, from Belgium having a very comprehensive list including employment status and urbanization, to Israel risk adjusting only on age \citep{Vandeven2007}. The authors found evidence of increasing risk selection in all five countries, including aggressive commercial campaigns and selective advertising in Belgium, Germany, and Israel and delayed reimbursement in Switzerland \citep{Vandeven2007}. 

Although \cite{Vandeven2007} claim that well-implemented risk adjustment is an ``essential precondition for reaping the benefits of a competitive health insurance market,'' it is clear that risk adjustment has historically led to suboptimal instances of risk selection as shown both inside and outside of the United States.

\section{Affordable Care Act Insurer Risk Adjustment Dataset Documentation}
\label{appendix:dataset_documentation}

\subsection{Brief Description}
For each year, the Affordable Care Act Insurer Risk Adjustment Dataset provides a combination of all "Insurer Report" Excel files and "Summary Report" PDF files. The "Insurer Report" Excel files include companies' basic information and insurer financial data; the "Summary Report" PDF files include accurate risk adjustment and reinsurance values.

\subsection{Dates of Data Collection}
2014, 2015, 2016, and 2017 Affordable Care Act Benefit Years.

\subsection{Data Source} The publicly-available insurer data used was downloaded from the CMS website. The 2014 and 2015 benefit year data was downloaded in late 2017. The 2016 benefit year data was downloaded in early 2018. The 2017 benefit year data was downloaded in early 2019. The 2014 and 2015 data has not changed since its downloading in late 2017. The datasets were created for 2014, 2015, 2016, and 2017. 

The Insurer Report data was downloaded from \href{https://www.cms.gov/apps/mlr/mlr-search.aspx}{https://www.cms.gov/apps/mlr/mlr-search.aspx}. The Summary Report data was downloaded from \href{https://www.cms.gov/CCIIO/Programs-and-Initiatives/Premium-Stabilization-Programs/index.html}{https://www.cms.gov/CCIIO/Programs-and-Initiatives/Premium-Stabilization-Programs/index.html}.

\subsection{Data Attributes} The data includes:

\begin{enumerate}
  \item "Insurer Report" "Company Information" Sheet - including Company Name, HIOS ID, Nonprofit Status, State
  \item "Insurer Report" "Summary of Data" Sheet - including Premiums, Claims, Taxes, Healthcare Quality Improvement Expenses, Non-Claims Costs, Income from Uninsured Plans' Fees, Amount Insured (Member Months, Number of Covered Lives, etc.), and Net Investment Income
  \item "Insurer Report" "Premiums and Claims" Sheet - including more detailed Premiums and Claims
  \item "Summary Report" Risk Adjustment Data - including Individual Group Reinsurance and Risk Transfers, and Small Group Risk Transfers
\end{enumerate}

\subsection{Data Format} All files are in the CSV format. Each row represents a company operating in a state. Each column represents a quantity, such as risk adjustment in the individual group, or member months in the small group market. The "Summary Report" and "Insurer Report" files were merged using an outer join on "HIOS ID," the 5-digit Health Insurance Oversight System Identification number for each company.

The names for columns corresponding to data from the "Summary Report" PDF Files are the names as they appear on the file: "HIOS ID," "HIOS INPUTTED INSURANCE COMPANY NAME," "STATE," "REINSURANCE PAYMENT AMOUNT (OR NOT ELIGIBLE)," "HHS RISK ADJUSTMENT TRANSFER AMOUNT (INDIVIDUAL MARKET, INCLUDING CATASTROPHIC)," "HHS RISK ADJUSTMENT TRANSFERS AMOUNT (SMALL GROUP MARKET)." These comprise the first six columns in our dataset.

The names for columns corresponding to data from the "Insurer Report" Excel Files are concatenated row and column names from that file without spaces. For example, if the "Insurer Report" row was "1.1 Direct premium written" and column was "2 Health Insurance INDIVIDUAL Total as of 3/31/17", then the column name in our dataset would be "1.1Directpremiumwritten2HealthInsuranceINDIVIDUALTotalasof3/31/17". These comprise the remaining columns in the dataset.

\subsection{Data Limitations} This dataset is limited by the number of companies that decided to report their estimated risk adjustment values. While the original PDF data sums to zero for every state, when matched with the Excel data, the actual transfer payments do not sum to zero.

\section{Testing for normality in risk transfers without assuming a linear cost term}
\label{Appendix:additional_tests}

In this section, we present two calculations of the probability that empirical $\overline{T_i}$ are distributed the way they are given that $\overline{T_i}$ follow the normal distribution $N(0, \beta^2)$. Unlike the strategies given in Section \ref{section:estimation}, these calculations do not make any specific assumptions about the kind of deviation from normality. Specifically, we do not use the assumption that there exists an additional cost term that scales with $\sqrt{n}$ as given in Equation \ref{eq:hypothesis} (i.e., insurers are not facing ``linear drift'' within their patient samples).

We run these calculations only on data from specific competitive state-years (CA, WI, and NY in 2014-2017). This is because even though we normalized dollar values to be in average 2017 dollars for all states, states and years have vastly different landscapes for the magnitudes of transfers due to the differing structures of the markets. For example, we wanted to exclude state-years that are less competitive, perhaps with one insurer dominating the market, because they may have lopsided transfers by nature of being less competitive. 

\subsection{Shapiro-Wilk Test}

The Shapiro-Wilk (SW) test on $\overline{T}$ tests the null hypothesis that a sample $\overline{T_1}, \overline{T_2}, ..., \overline{T_n}$ comes from a normally distributed population \citep{ShapiroWilk}. The test statistic $W$ is a ratio roughly representing how far the empirical sample deviates from a normally distributed sample, and the statistic p-value indicates rejection of the null hypothesis \citep{ShapiroWilk}. 

\subsection{Normal Comparison Test}

Similar to the Shapiro-Wilk Test, this test tries to reject normality by comparing our empirical $\overline{T_i}$ to a normal distribution with some standard deviation. Say we find too many points in the exterior few standard deviations and too many points in the interior close to the mean, and the probability this occurs is statistically impossible. Then, we can reject the null hypothesis that the empirical distribution comes from a population that is normally distributed. We call this test the Normal Comparison (NC) Test.

We want to make a comparison between our absolute transfers per member month $|\overline{T_i}|$ with some $\Delta_0 \in X' \sim N(0, 1)$. First, we define $a_0$ as the point on $X \sim N(0, 0.5)$ that we have $P(X>a_0) = \frac{1}{2}\frac{p}{k}$. Then $a_0 = erf^{-1}(erf(a_0)) = erf^{-1}(P(X\in [-a_0, a_0])) = erf^{-1}(1-2P(X > a_0)) = erf^{-1}(1-\frac{p}{k})$. Then, we transform $\Delta_0 \in X' \sim N(0, 1)$ to be in terms of $a_0$: $\Delta_0 = \sigma\sqrt{2}a_0 = \sqrt{2}erf^{-1}(1-\frac{p}{k})$. $\Delta_0$ represents the distance from $0$ such that $P(X'>\Delta_0)=\frac{1}{2}\frac{p}{k}$ in $N(0,1)$. Because $\sigma=1$ in $N(0,1)$, $\Delta_0$ also represents the number of standard deviations away from the point where $P(X'>\Delta_0) = \frac{1}{2}\frac{p}{k}$. 

Define $\beta_0 = \frac{max |d|}{\Delta_0}$. $\beta_0$ is the smallest possible value we will accept for $\beta$, because it scales the maximum $|\overline{T_i}|$ point to be on $N(0, 1)$ where the probability that we are in the tail to the right of this point is $\frac{1}{2}\frac{p}{k}$. To show significance for this test, we must consider two cases:
\begin{itemize}
    \item \textbf{Case 1 (``one point too far from the mean''):} In this case, we want to show that the dataset an unusually large deviation from the mean for the distribution to be normal. In this situation, our dataset's $\beta_0$ would be greater than the population's $\beta$. We calculate $\Delta_0$, which is the ratio needed for the probability of seeing the maximum $|\overline{T_i}|$ among $k$ draws to be $<\frac{p}{2}=0.025$. This means for any $\beta' < \beta_0$, there is one point that is significantly far away from the mean, so the distribution cannot be normal. 
    \item \textbf{Case 2 (``too many points in the interior''):} \label{case2} In this case, we want to show that the dataset has far too many points within 0.1 standard deviations of the mean for the distribution to be normal. We choose 0.1 standard deviations because 1 standard deviation is not a sufficiently rare event to show significance. In this situation, our dataset's $\beta_0$ would be less than the population's $\beta$. We count for how many $i$, $|\overline{T_i}| < 0.1 \beta_0 < 0.1 \beta$ and label this count $s_2$. 
\end{itemize}

We calculate the probability $p_{0.1}$ of having an $i$ for which $|T_i| < 0.1 \beta_0 < 0.1\beta$, or in words whether the point $|T_i|$ is within 0.1 standard deviations of the mean:
\begin{align*}
p_{0.1} &= P\left(\overline{T}\in [-0.1\beta_0, 0.1\beta_0]\right) < P\left(\overline{T}\in [-0.1\beta, 0.1\beta]\right) \\
&= P\left(X'\in[-0.1, 0.1]\right) = erf\left(\frac{0.1}{\sqrt{2}}\right)\approx 0.079655
\end{align*}

Then, we calculate the probability $p_2$ of $s_2$ or more successes, each of probability $p_{0.1}=0.0797$, in a sample size of $k$, using the binomial random variable $B_2 \sim Binom(k, p_{0.1})$: 
\begin{align*}
    p_2 &= P(B_2 > s_2) = 1 - P(B_2 \leq s_2) \\
    &= \sum_{i=s_2+1}^{k}\binom{k}{i}p_{0.1}^{i}(1-p_{0.1})^{k-i}
\end{align*}

For significance of this NC test, we need $p_2<\frac{p}{2}=0.025$. 

\begin{table}[h]
\centering
\begin{tabular}{|c|c|c|c|c|}
\hline
\textbf{State} & \textbf{Year} & \textbf{SW W-statistic} & \textbf{SW p-value} & \textbf{NC p-value} \\ \hline
CA only & 2014 & 0.8497 & 0.022* & 0.687 \\ \hline
CA only & 2015 & 0.9567 & 0.6017 & 0.735 \\ \hline
CA only & 2016 & 0.9375 & 0.3198 & 0.368 \\ \hline
CA only & 2017 & 0.9493 & 0.5138 & 0.027* \\ \hline
WI only & 2014 & 0.7147 & 0* & 0* \\ \hline
WI only & 2015 & 0.8352 & 0.0004* & 0.007* \\ \hline
WI only & 2016 & 0.9627 & 0.4484 & 0.052 \\ \hline
WI only & 2017 & 0.9719 & 0.7344 & 0.276 \\ \hline
NY only & 2014 & 0.8868 & 0.0235* & 0.001* \\ \hline
NY only & 2015 & 0.7842 & 0.0005* & 0.001* \\ \hline
NY only & 2016 & 0.8418 & 0.005* & 0.059 \\ \hline
NY only & 2017 & 0.8103 & 0.0012* & 0.481 \\ \hline
\end{tabular}
\caption{Results of the SW and NC Tests}
\label{tab:nctesttable}
\end{table}

\subsection{Results}

Our comparison of the $|\overline{T_i}|$ to a normal distribution showed significant results only for NY and WI in 2014 and 2015, as seen in Table \ref{tab:nctesttable}. Based on the results of this test, we looked to creating a stronger test based on our hypothesis that the empirical distribution stochastically dominates the theoretical distribution to some degree of statistical significance at at least one point. We use stronger assumptions than the simple non-normality we wanted to show by this first test in creating other empirical strategies.

\section{Supplemental Tables}
\label{Appendix:supplemental_results}

\begin{table}[htbp]
\centering
\begin{tabular}{|c|c|c|c|c|}
\hline
\multicolumn{2}{|c|}{\textbf{Comparison to}} & \multicolumn{3}{c|}{\textbf{Simulation described in Alg.  \ref{alg:simulated_transfers}}} \\ \hline
\textbf{State(s)} & \textbf{$k$} & \textbf{$\beta$} & \textbf{$P(|\overline{T'_i}|) >2\beta$} & \textbf{p-value} \\ \hline
Random A & 976 & 17,540 & 0.178 & $<10^{-6}$ \\ \hline
Random B & 990 & 19,280 & 0.180 & $<10^{-6}$ \\ \hline
All & 1966 & 16,700 & 0.198 & $<10^{-6}$ \\ \hline
\end{tabular}

\caption{Supplemental Results for estimated $\beta$ from comparing the Empirical $|\overline{T}|$ against $|\mathcal{N}(0,\beta^2)|$ and simulated $|\overline{T'}|$.}
\label{tab:beta_estimation_results_Appendix}
\end{table}

\begin{table}[]
\centering
\begin{tabular}{|c|c|c|c|}
\hline
\textbf{State(s)} & Realized $\sum_{i\in I} |T_i|$ & $E(\sum_{i\in I} |T'_i|)$* & $E(\sum_{i\in I} |T'_i|)$** \\ \hline
NY only & $1.83\cdot 10^9$ & $1.04\cdot 10^9$ & $9.17\cdot 10^8$ \\ \hline
CA only & $2.04\cdot 10^9$ & $1.03\cdot 10^9$ & $1.11\cdot 10^9$ \\ \hline
WI only & $1.30\cdot 10^8$ & $9.04\cdot 10^7$ & $9.87\cdot 10^7$ \\ \hline
Competitive & $7.36\cdot 10^9$ & $2.63\cdot 10^9$ & $2.65\cdot 10^9$ \\ \hline
Random States A & $4.00\cdot 10^9$ & $1.43\cdot 10^9$ & $1.33\cdot 10^9$ \\ \hline
Random States B & $4.62\cdot 10^9$ & $1.59\cdot 10^9$ & $1.62\cdot 10^9$ \\ \hline
All & $8.62\cdot 10^9$ & $2.74\cdot 10^9$ & $2.75\cdot 10^9$ \\ \hline
\end{tabular}\\
\caption{The realized empirical and expected simulated absolute transfers in average state 2017 dollars.}
* from Alg. \ref{alg:simulated_transfers}
** from percentile simulations
\label{tab:expected_transfers}
\end{table}

\clearpage

\section{Supplemental Figures}
\label{Appendix:supplemental_results_2}

\begin{figure}[h]
\begin{center}
\includegraphics[height=3in]{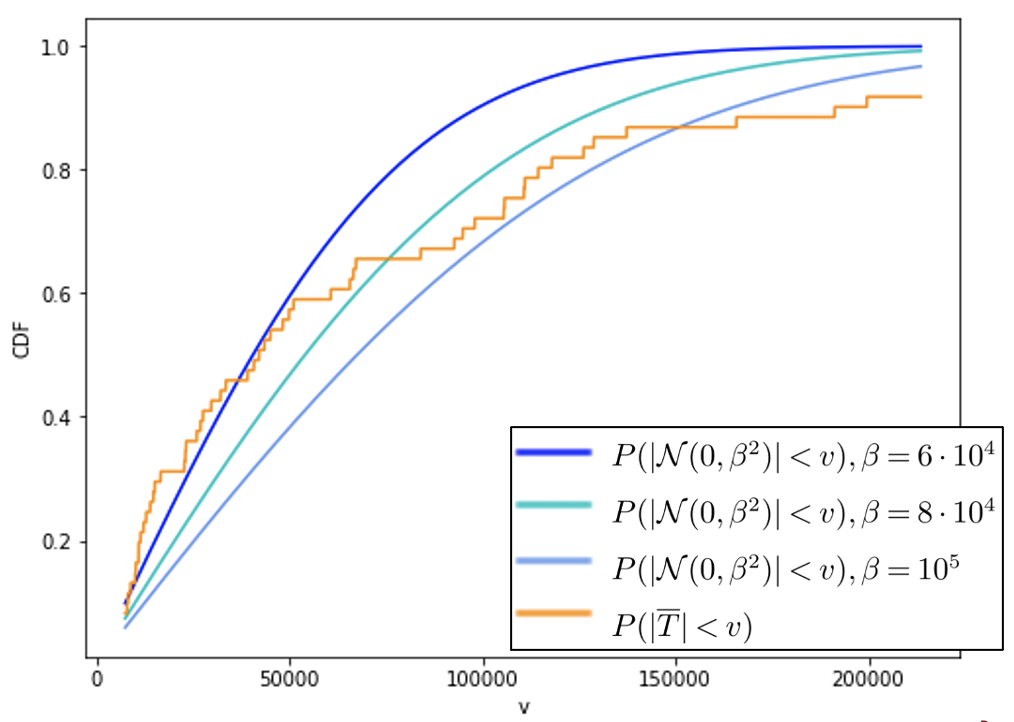}
\caption{An example of estimating $\beta$ from California with $k=61$ insurers over 4 years. The CDF of the empirical distribution is shown in orange. Half-normal distributions are shown in blue, with various values for $\beta$ of $60,000$ on the left, $80,000$ in the center, and $100,000$ on the right. Approximately up to $v=40,000$, $|\overline{T}|$ dominates $|\mathcal{N}(0,\beta^2)|$. The point of maximum difference for each $\beta$ is the value $v$ for which there is the greatest difference between the empirical and half-normal distributions.} \label{fig:estimating_beta_example}
\end{center}
\end{figure}

\begin{figure}[h]
\begin{center}
\includegraphics[height=3in]{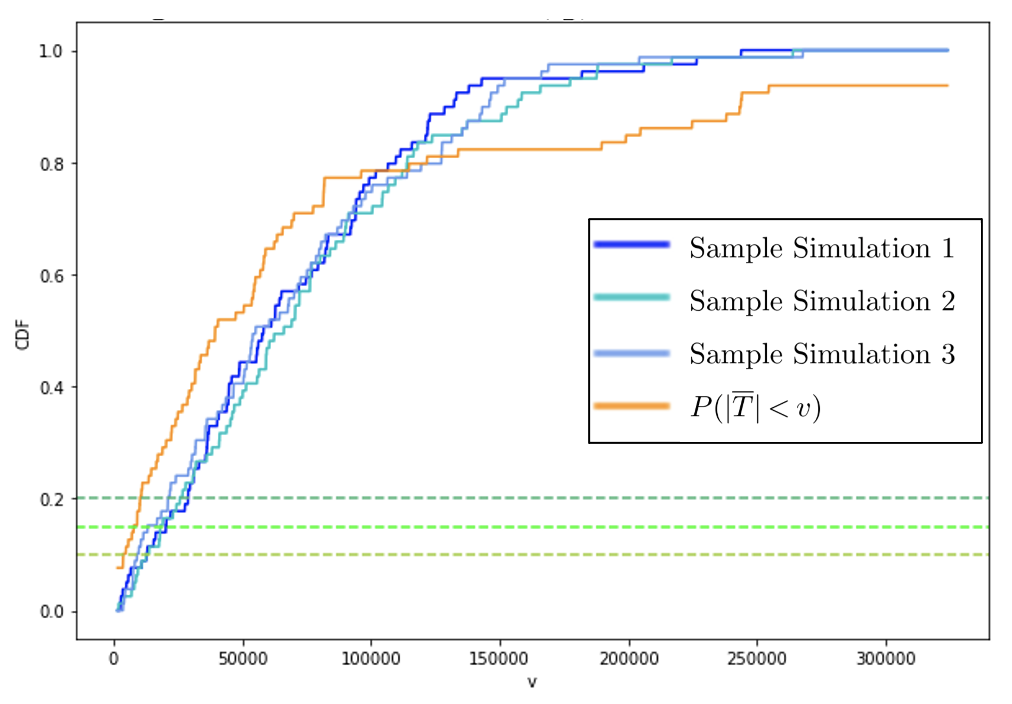}
\caption{An example of estimating $\beta$ from New York with $k=79$ insurers over 4 years. The CDF of the empirical distribution is shown in orange. Given a fixed $\beta$, we generate $10^5$ sample simulations, three of which are shown in blue. At each of the percentiles $p=\{0.1, 0.15, 0.2\}$ shown as horizontal green lines, we find the $\beta$ for which the empirical distribution is at most in the leftmost 1\% of simulations generated.} \label{fig:estimating_beta_example_2}
\end{center}
\end{figure}

\end{document}